# A closer look at Intrusion Detection System for web applications

Nancy Agarwal [1*], Syed Zeeshan Hussain [2]

[1] Department of Computer Science, Jamia Millia Islamia, New Delhi, India
[2] Department of Computer Science, Jamia Millia Islamia, New Delhi, India
*nancy.agarwal02@yahoo.in
[2]szhussain@jmi.ac.in

**Abstract:** Intrusion Detection System (IDS) acts as a defensive tool to detect the security attacks on the web. IDS is a known methodology for detecting network-based attacks but is still immature in monitoring and identifying web-based application attacks. The objective of this research paper is to present a design methodology for efficient IDS with respect to web applications. In this paper, we present several specific aspects which make it challenging for an IDS to monitor and detect web attacks. The article also provides a comprehensive overview of the existing detection systems exclusively designed to observe web traffic. Furthermore, we identify various dimensions for comparing the IDS from different perspectives based on their design and functionalities. We also propose a conceptual framework of a web IDS with a prevention mechanism to offer systematic guidance for the implementation of the system. We compare its features with five existing detection systems, namely AppSensor, PHPIDS, ModSecurity, Shadow Daemon and AQTRONIX WebKnight. This paper will highly facilitate the interest groups with the cutting edge information to understand the stronger and weaker sections of the domain and provide a firm foundation for developing an intelligent and efficient system.

1. Introduction

There has been a phenomenal rise in the use of web-based applications over the last decade. Applications such as e-banking, e-commerce, online blogs, social networking sites, etc. have become a common platform for transmitting information and delivering online services. Although web applications offer great digital experiences, only the secured ones can deliver the services safely. Since these applications deal with sensitive data and operations, for attackers they are an easy, lucrative and potential target to acquire confidential data, earn monetary gain and perform several unlawful activities [40, 36]. Nowadays, web application security is one of the pertinent issues in information security due to the continuous growth in the number of web attacks. According to the Internet Security Threat Report (ISTR) 2017 [23], more than 76% of the scanned were found vulnerable. A survey reports that 60% of the hacker attacks either targeted the web applications or utilized them as the attack vectors. The latest report by Verizon [1] says 95% of the web application breaches are financially motivated[1]. Moreover, as per the report [87], the number of web-based security breaches in the first quarter of the year, 2017 has been increased by 35% from the previous year.

Several defense mechanisms have been adopted by organizations to ensure adequate security for web applications. Web Application Firewall (WAF) [8] is the most prominent defense mechanism used by the organizations to protect the web application after the deployment.WAF first analyzes the web requests before they are sent to the web application and blocked if found malicious. It is, however, a sort of a generic product that operates by rules, and protects commonly known attack sequences but is insufficient to understand the context of custom web applications. Configuring WAF is like designing an alarm system for a building without considering the blueprint of the building. IDS, primarily designed to protect web applications is as a potential solution. IDS is the defense tool which is used to detect and report the intrusions to the administrator. IPS (Intrusion Prevention System) is the extension of IDS which is capable of responding to attack incidents and block if necessary. IPS works similar to WAF, but it can inspect in-depth traffic details to detect an attack. These tools offer a more dynamic solution than WAF as they can utilize both signature-based technique and anomaly-based technique for detecting known threats and identifying abnormal behavior respectively. They can learn the complex ecosystem of a web application such as how business logic works, who are the users, how they interact with the application and so on.

---

[1] WhiteHat Security Application Security Statistics Report. https://info.whitehatsec.com/rs/675-YBI-74/images/WHS%202017%20Application%20Security%20Report%20FINAL.pdf.

Intrusion detection methodology is still new in the domain of web application security. IDS are primarily designed to observe and detect intrusive activities on the network [27]. However, the characteristics of network-based attacks are significantly different from the web-based attacks. The former targets the network layer while the latter focuses on the application layer. Secondly, current web applications are complicated, database driven and usually created by developers with limited security skills. These applications are highly customized, provide dynamic content, facilitate interactive user sessions and conduct sophisticated business operations [30]. The attack surface varies with the respective business logic and skill set used to design these applications. Therefore, creating an IDS for recognizing suspicious activities on a website requires a significantly different approach than the IDS designed to monitor network traffic.

The objective of the paper is to thoroughly understand and characterize the design methodology of the detection system exclusively built to monitor web traffic. In this paper, we discuss a number of unique characteristics of the web applications and its traffic which pose challenges to designing a web IDS and explain their effects concerning the design of IDS. It would highly facilitate the developers to craft an efficient architecture of the web IDS. The paper also presents a literature survey of research on IDS, proposed as a solution to web application security. It aims to explore the measures which have been suggested in existing studies to deal with the issues of web application security and identify the current state of the art and future perspectives. Based on the literature review, we recognized nine dimensions of detection systems to compare the existing systems. IDS tools are generally evaluated on performance metrics such as detection coverage and false-positive rates [94]. The identified dimensions assist in comparing the existing web IDS in terms of their design methodology and functionalities. The comparison provides useful information about the current state of IDS in web application security and several exciting trends and challenges of the domain. Besides, we also offer a conceptual framework of intrusion detection system with prevention mechanism (IPS) that offers systematic guidance of the design and implementation of a system for web applications. We compare its features with 5 well-known detection systems, namely AppSensor, PHPIDS, ModSecurity, Shadow Daemon and AQTRONIX WebKnight which highlight the strength and additional functionalities of the proposed framework.

The rest of the paper is structured as follows. Section 2 describes major approaches for detecting intrusions. Section 3 highlights several features of web applications and its traffic which are highly responsible for the performance of IDS. Section 4 provides a glimpse of security measures offered by IDS to prevent web-based attacks. Section 5 throws light on a comprehensive literature survey conducted to explore various intrusion detection systems proposed by researchers for web applications. Section 6 introduces the nine identified dimensions of web IDS for comparing the existing systems. In section 7, the current state and associated challenges have been highlighted in the form of a graph. Section 8 explains the conceptual framework of web IPS and compare it with the five other detection systems. Finally, section 9 concludes the paper followed by a brief outline on the future direction in section 10.

## 2. Intrusion Detection Approaches

There are major four approaches towards intrusion detection, namely misuse, anomaly, policy and hybrid. Each detection approach works on a specific set of principles. The current section briefly states the mechanism of each detection technique along with the challenges faced by them.

### 2.1. Misuse-based Intrusion Detection

The misuse-based approach uses a set of signatures representing the patterns of already known attacks to filter malicious activities. Misuse-based systems (also known as signature-based systems) have the capability of detecting known attacks more precisely with less false positive rate but prove to be inefficient for detecting zero-day or unknown ones. The signature database should be up to date for recognizing novel attacks which is quite tedious and intensive process since new attacking techniques are being frequently discovered [23]. Another issue arises because of the vulnerabilities caused due to the in-house developed web applications. Such vulnerabilities demand an expert knowledge for creating signatures specific to custom vulnerabilities other than common vulnerabilities. Next challenge arises due to the existence of infinite variants of attack vectors. A slight change in attack vector might easily fool the signature-based detector. Therefore, signatures should be general enough to cover these variants. However, there exists a trade-off between deciding their specificity and sensitivity level. Sensitive (general) signatures amplify the risk of getting high false positive alerts while selective (specific) ones fail in detecting attack variants.

### 2.2. Anomaly-based Intrusion Detection

The anomaly-based technique assumes that malicious activities are significantly different from expected behavior, and that can be studied quantitatively. The incoming events are analyzed to check whether they deviate from the normal ones.

Unlike misuse, anomaly-based systems support detection of unknown and novel attacks and can also be trained to cater the problems caused by the custom vulnerabilities. Besides having great potential, there are some critical issues associated with the approach. Anomaly-based components can model the acceptable behavior by using the multitude of different machine learning techniques [37], and selecting best ones is a significant issue. Also, deciding the optimal thresholds of the machine learning parameters is challenging. The high threshold value may increase the number of undetected attacks whereas too lenient configuration may cause higher false positive alerts. However, the solution depends upon the variability of the behavior being observed. If the web traffic is highly variable, the generalized model can handle the situation better while low variable traffic requires a strict model to detect doubtful movement. Anomaly-based systems are also known for producing high false positive rates which may cause blocking or denying of a good number of legitimate requests. The assumption (i.e., attacks manifest unusual behavior) behind the approach is the prime reason since at times a benign user does exhibit strange behavior which might not have been recorded during the training phase.

*2.3. Policy-based Intrusion Detection*

The two discussed techniques suffer from their inherent limitations. Misuse detection can never maintain data of all possible attack vectors, and likewise, anomaly detection cannot record all legitimate behaviors of users. An alternative approach, namely policy-based intrusion detection is receiving considerable attention these days as it allows to overcome these limitations. Policy-based techniques establish boundaries between the allowed and not allowed events by imposing a set of rules [26]. It solves two major problems: (1) detection of unknown attacks, (2) classification of normal unseen behavior into attack class. Although this approach seems useful and flexible but there are also certain drawbacks associated with it. First, a security specialist is required to design effective policies. Second, defined policies should be consistent and in a logically correct state throughout the system to avoid any adverse circumstances. Policies are inter-related through their associated conditions, and therefore, there may exist inter or intra-policy conflicts as an incoming event may trigger more than one rule either within a policy or between two policies. Moreover, these policies are usually implemented sequentially, and improper ordering can cause a feedback loop or deadlock situation. However, ontology-based systems [3-5] can be used to simplify the policy specification and management tasks.

*2.4. Hybrid Intrusion Detection*

A hybrid system is the fusion of different intrusion detection approaches into a single integrated detection system [38, 45]. Hybrid based systems give better performance by utilizing the strength of more than one approach to overcome the limitations of individual techniques. However, while incorporating the different methods, a few things should be taken into consideration. First, hybrid systems can have either a layered or parallel architecture but opting one of them is a preliminary requirement. Moreover, in layered architecture deciding the correct sequence of multiple components for processing events is another challenge. For example, the authors in work [45] proposed the hybrid system where the anomaly detection component is placed first followed by the misuse component. The second point to be considered is how to resolve the conflicts between results classified by these components since there may be the case when one classifies an event into a safe class, and other declares the same as an intrusive [46].

Each of the detection techniques has both strengths and weaknesses. Misuse detection technique generates less false-positive alarms but may miss an attack if the respective pattern is not stored in the database. Similarly, the anomaly-based mechanism is capable enough to recognize novel attacks but causes high false-positive rates as it is inefficient in differentiating the malicious from unknown legitimate behavior. Logically, the misuse-based approach uses knowledge of malicious behavior (known attacks), and anomaly-based approach uses knowledge of normal behavior. Meanwhile, policy-based approach tries to set equilibrium between normal and malicious behavior via enforcing the set of rules. It therefore seems to be a more viable detection system but inter- and intra-policy relationships cumbersome the whole policy management task. And the last detection technique, the hybrid approach is aimed towards utilizing the potential of multiple methods in such a way that the strength of other overcomes the weakness of one. However, the fact that how well the different components are integrated determines the performance of hybrid-based systems.

In the next section, we are discussing the peculiarities of web applications and its traffic which profoundly affect the performance of IDS. Knowledge of these features would significantly help the upcoming researchers to develop a sound understanding of the detection system and design an efficient framework.

## 3. Challenges in Intrusion detection for web applications

Intrusion detection methodology is still immature in web application security domain. The detection systems are primarily used as a network security appliance. However, designing the web IDS requires a different approach from the traditional network IDS for handling the complexities associated with web-based applications. In this section, we discuss several characteristics of web application and web traffic which make the process of developing the IDS a challenging task. The characteristics explained in the following subsections underlie the theoretical basis for designing the web IDS and will assist in having a thorough grasp of the basic knowledge required to create a robust architecture of the system.

*3.1. Communication Protocol (HTTP/HTTPS)*

Attackers exclusively make use of HTTP/HTTPS protocols to exploit the web application vulnerabilities. Hypertext Transfer Protocol (HTTP) [9] is a request-response protocol designed to facilitate communication between the client and server, and HTTPS [39] ensures the secure and encrypted connection. When HTTPS traffic is observed from the IDS perspective, one significant disadvantage is that encryption blindfolds the network-based detection systems. IDS can be characterized into Host-Based Intrusion Detection System (HIDS) and Network-Based Intrusion Detection System (NIDS) on the basis of whether they operate on an Application layer or Internet layer of TCP/IP model [20]. NIDS monitors the network packets, and in HTTPS communication, the packet data exists in the encrypted form which fails the system to inspect it. However, these systems can check HTTPS traffic if they have access to the private key of the SSL certificate. On the contrary, HIDS does not face any problem in dealing with HTTPS traffic as they protect endpoints where encrypted information is decrypted back into the normal form.

*3.2. Web Request*

Web request carries data from the client to the server. The data is passed using either HTTP request header fields or request parameters [34]. Request header fields contain control information of the client request while the request parameters include additional user information required by server-side programs to conduct an operation. GET and POST are the two standard methods used to pass parameters to the server [34]. In the GET request, parameter values are passed in the query string of URL, and in the POST request, these values are carried in the request body. Client browser generally sets header fields whereas the parameter values are either provided by the user or were already initialized by server-side programs (e.g., cookies, hidden fields, etc.). The fundamental problem with web-based application security is, the data provided by clients can be highly variable and equally complex, and therefore it becomes complicated to bind them by a valid set of values.

The primary role of detection systems revolves around observing the values provided in the header fields and request parameters. These values can be validated using either positive or negative approach. Positive validation approach defines what data is expected by the application. It consists of several types of validation checks such as data type (string, integer), minimum and maximum length, specific patterns, etc. On the other hand, the negative approach involves filtration of values which contain the attack patterns. Ontology and signature-based systems include both positive (white-listing) and negative (black-listing) validation while the anomaly based systems are concerned with positive validations only. The data passed in a web request consists of a variety of values, and the choice of selecting an appropriate approach (white-list or black-list) depends significantly on the type of value set. We have categorized the value set into following classes.

*3.2.1 Finite Values:* These values exist in a finite range and can be either independent of application's business logic (i.e., common to all) or specific to it. Former group contains a set of common values, for example, header fields like Accept, Accept-Charest, Accept-Language, etc. These values are mostly the same for different applications, and so can be validated against a general white-list in SIDS. The latter category includes parameters which contain values of HTML controls like dropdown lists, check-boxes, etc. These controls are used to assist clients to choose values from the given set only. The value set of these parameters depends on the business logic of an application. Maintaining the white-list to validate such parameter values can be turned into a tedious task for SIDS due to many reasons. Primarily, the whitelist is no longer the general enough but rather, is specific to the set of values corresponds to the business logic. Second, this list can be reasonably long depending upon the number of controls in an application. Third, maintenance of the list is a complicated process since allowed set of values may get modified promptly with the change in the business logic. However, AIDS can work well in this scenario since it is capable of learning the values of the parameters automatically [53].

*3.2.2 Fixed format values:* As the name suggests, these values follow a specific format, e.g., parameters which carry values like date, time, etc. To analyze these values, one can make use of either AIDS by learning the character distribution and structure of these values [53] or whitelist of SIDS by defining the pattern of allowed values.

*3.2.3 Application values:* This class represents the values which are given by server-side programs and must not be tampered at the client side. Developers store a variety of essential information like price and quantity of products, session id, etc. in the form of cookies, hidden fields or query strings. IDS need to ensure that these values are same as set by the application. Signature-based IDS seem to be inefficient in recognizing tampered values as these systems demand an attack pattern, and tampered values usually appear as similar to a normal data. However, anomaly-based systems can be used to learn the parameters whose values should not be altered at the client-side. The work presented in [49] encountered the parameter tampering attacks.

*3.2.4 Infinite and complex values:* The parameters such as username and password in a login form or user comments on a blog carry plaintext data. These parameters mostly have an arbitrary set of values, and therefore, are generally validated against known attack patterns specified in a signature blacklist.

*3.3. Multiple users with multiple roles*

Web applications are usually accessed by multiple users with different level of privileges [29], and these privileges are controlled by the authorization process which ensures that the user is performing only permissible operations. Applications use session management mechanism [28] to track an individual client-server interaction and map the request to a particular user for deciding whether the request should be processed or denied. A session ID is associated whenever a user login to the application for distinguishing the series of requests from the request pool and link them to the logged in user.

Facilitating different users with a different set of privileges imposes several demands on the detection systems. Primarily, IDS must have the functionality to track user sessions to link the client requests to their specified session. Second, the system should also have the capability to monitor the integrity of a session for ensuring whether the session is being used by the same person who logged into the system. There are various ways to take over sessions illegitimately like session hijacking and session fixation [28]. The third demand says IDS should also perform continuous monitoring of resource usage and user activities within a session. A well-formulated privilege escalation attack can lead to the unauthorized access, which may even allow the attacker to gain administrator privileges. "Stateful Analysis" is the necessary feature required by any IDS to satisfy these demands. This feature assists the IDS to track the state of an individual session. While stateless IDS treats every request independent of each other and do not maintain any track of requests [46], [47], the stateful mechanism can correlate the sequence of requests to a specific user [48], [50]. Systems which do not possess any mechanism to map the current request with previously arrived requests are unlikely to recognize exploitation against state maintenance and authorization.

*3.4. Continuous Changes*

Like any software development life cycle, web applications also need maintenance and demand continuous modifications for adding new functionalities, absorbing new technologies, etc [100]. Depending upon the application domain, these changes may be either frequent or regular. Applications like e-commerce and social networking sites continuously evolve to enhance user experience and engage more users towards them[2], whereas those web applications which are basically designed to mark the organization presence online rarely change.

The continuous changes made to an application over time introduce a major challenge to IDS. The detection systems also need to be tuned and maintained with time to accommodate changes made in an application. In the context of anomaly detector, frequent changes handicap the system in a sense that the present model is unaware of the modified version of the application. Apparently, it would misclassify new legitimate behavior as an intrusive. These detectors require retraining to adapt the changes [42]. Likewise in the case of policy-based systems, the rules should be re-defined to incorporate the changes. However, in the signature-based system, the blacklist is not much affected since attack patterns remain the same, but the whitelist needs to be updated with respect to the changed behavior of the application.

*3.5. Dynamicity*

The web application consists of a mixture of static and dynamic web pages. Static implies constant, a web page that presents the exact contents before users as it is stored at the server side. Whereas dynamic pages include web pages whose content is generated on-the-fly by server-side or client-side languages, mostly based on the parameter values passed in an HTTP request. Dynamicity profoundly affects the performance of intrusion detection systems. For static pages, there exists a deterministic relationship between the user requests and the content to be displayed on the browser [13]. Requests for static pages do not require any specific data from the user which significantly simplifies the task of IDS. On the other hand, dealing with dynamic web pages is the main challenge faced by intrusion detection systems. Dynamic web services usually need parameters from client side for processing, and this feature is highly exploited by attackers to hack into a website. Dynamicity complicates the job of IDS, the more the dynamicity the more is challenging for IDS to detect intruders.

---

[2] http://www.techtimes.com/articles/136865/20160229/guide-latest-facebook-changes-use-new-features.htm

*3.6. Heterogeneity*

The web application can be implemented using various server-side languages, such as PHP, Java, Perl, etc. NIDS are independent of languages, and hence same IDS can be deployed for different server-side language applications. On the other hand, HIDS is language specific and can be designed either general or specific to a particular language. For example, Snort is an open source NIDS used to detect web-based attacks for any application [33], and PerlIDS is a host-based IDS used to protect web applications implemented in the Perl language[3].

*3.7. Bot Requests*

The web traffic can also be generated by bots rather than humans. Bots are automated scripts which are designed to execute a set of specific activities on websites. These scripts can be of two types. First, scripts which are designed to mimic certain human behavior [95] like automating clicks on online ads (click fraud), submitting forms, flooding the websites with millions of requests to perform Denial of Service (DoS) attacks, etc. Second, the scripts which are designed to launch web attacks such as Angler, RIG and Neutrino attack toolkits [23], [32]. Scripts can perform tasks faster than manual without requiring much human involvement. Existing attack toolkits provide several additional benefits over manual attacks. These toolkits assemble several attack techniques in a single package and assist the novice attackers with the ability to mount attacks on websites in a relatively more straightforward way. The attacker can use the toolkit to exploit the vulnerabilities without in-depth technical knowledge. Furthermore, automated attacks are comparatively cheaper than manual attacks since they allow adversaries to target a large number of websites with less time and effort [31].

Signature-based systems are not sound enough to recognize requests from the scripts designed to mimic the user activities because malicious requests differ from the legitimate ones in intent but not in content, and SIDS operates on malicious content rather than intent. However, the adversary can be caught using anomaly-based IDS from several abnormal behavioral features. For example, automated tools usually generate web requests at a higher frequency than the regular rate of requests, and this extreme change in request behavior can easily be recognized by AIDS [53]. On the other hand, automated attacks can be best handled by SIDS since the scripts use common attack vectors to target the vulnerabilities [59].

In the section, we discussed several particular features of web application and its traffic that would play a crucial role while designing the framework of an IDS. A quick snapshot of the impact of features on IDS is shown in Table 1.

**Table 1** Challenges of web IDS

| Characteristics of web application and its traffic | Impact on IDS |
|---|---|
| Communication Protocol (HTTP/HTTPS) | • HTTP communication is carried out in plain text whereas HTTPS communication is encrypted.<br>• NIDS fails in analyzing HTTPS traffic whereas HIDS can handle both HTTP and HTTPS. |
| Web request | • Web requests carry a variety of parameters.<br>• Detection approach highly relies on the type of the values of parameters. e.g.<br>    Business logic independent finite values: whitelist approach of SIDS<br>    Business logic dependent finite values: anomaly-based approach<br>    Fixed format values: the anomaly-based approach or whitelist approach of SIDS<br>    Application initialized values: the anomaly-based approach<br>    Parameters carrying text data: the blacklist approach of SIDS |
| Multiple users with multiple roles | • Web applications facilitate multiple user interactions through sessions and provide access rights based on user role.<br>• Stateless IDSs are not efficient enough to recognize attacks on session management and authorization policies.<br>• Stateful detection mechanism overcomes the limitation of stateless IDS as |

---

[3] http://search.cpan.org/dist/CGI-IDS/lib/CGI/IDS.pm

|  |  |
|---|---|
| | it can track and monitor an individual user session. |
| Continuous Change | • Continuous modifications in application source code directly affect the efficiency of IDS.<br>• AIDS requires retraining to accommodate the changes.<br>• Blacklist-based SIDS is not much affected. |
| Dynamicity | • The web application consists of static and dynamic web pages.<br>• The more the dynamic content, the more is the challenging task for IDS |
| Heterogeneity | • Web Applications can be implemented using different server-side languages.<br>• NIDS is independent of programming language whereas HIDS can be designed either general or specific to a particular language. |
| Automated Requests (Bots) | • Scripts can also be designed to issue HTTP requests automatically.<br>• SIDS suits best for detecting scripts designed to automate web-based attacks.<br>• AIDS suits best for detecting scripts designed to mimic human behavior. |

AIDS: Anomaly-based Intrusion Detection System       SIDS: Signature-based Intrusion Detection System

## 4. IDS Security

Web vulnerabilities are the weaknesses of a web application that allow a user to execute malicious operations. OWASP Top Ten Project [88] is dedicated to providing the list of prevalent and critical web application vulnerabilities. In order to secure web applications, intrusion detection systems provide various kinds of security measures to guard them against adversaries. Following subsections present a brief overview of different security features incorporated by IDS to identify the suspicious user activities.

### 4.1. Input Validation

The core problem with the insecure web applications is the fact that users have full control over the data to be submitted to the server. They can supply any arbitrary input into the parameters including header fields, or may even modify the values stored by the application at the client side in the form of hidden fields, cookies and URL parameters. A large proportion of applications rely on client side measures to control the user data, which, however, attackers can easily circumvent. From the security perspective, every piece of data received from the user should be treated as potentially malicious, and thoroughly verified at the server. Improper or no input validation at the server side is the root cause behind most of the web application vulnerabilities including XSS, SQL Injection, unvalidated redirects, etc. [25]. Input Validation is the key security feature provided by almost every detection system. Validation process performs checks if the input satisfies a predefined set of rules for preventing any entry of unsafe data into the application. It comprises checking all input parameters including URLs, form data, cookies and query strings.

### 4.2. Output Validation

HTTP response is generated as an output of the client request processed by the server. It contains a series of header fields and a message body. Generally, output validation guards an application against unintentional revealing of sensitive information. The application may reveal critical information in the form of cookies or hidden fields in an insecure way. For example, if "secure" flag in the cookie is not set by the application code while sending the credentials over HTTPS, the browser can transmit the credentials stored in the cookie over unencrypted HTTP channel. The application may also expose internal details if it does not handle the errors messages. For instance, error based exploitation is one of the SQL injection techniques used by a hacker to extract information from the database. There are detection systems which also verify the response data, and ensure that no potential harmful content is sent to the client in an insecure way. They mostly modify the content of the HTTP response to serve the purpose.

### 4.3. Session Verification

Almost every application requires some mechanism to associate incoming application requests to their specific users for facilitating client-server interaction in the form of sessions. Since HTTP is a stateless protocol, a Session ID (randomly generated alphanumeric string) is associated with each user request to keep track of his activities within a session. A large fraction of web attacks involves session hijacking, session fixation, session replay where an attacker steals or overwrites users' Session ID to impersonate the valid user and perform operations on his behalf [24, 84]. Session verification offered by IDS ensures that whether the session is being used by the same user who logged into the application. It may also include detecting any attempt being made by an adversary to get the session ID illegitimately.

*4.4. Access Control*

Access control defines the policies to regulate the privileges granted to the application user. Web applications serve multiple users with the different privileges which are usually given on the basis of their role (e.g., visitor, manager, admin, etc.). Some detection systems also provide the functionality of monitoring the activities of the clients to prevent any unauthorized access to the information or services offered by the application. The systems may also detect those requests that attempt to unauthorized access to the objects (e.g., files and directories) which are mistakenly exposed via URL or form parameter[4]. For example, an online publishing site may allow its users to purchase e-books by directing him to the download link once he makes the payment. In the case of improper access control verification, if the user already knows the link, he can access the resource without payment.

*4.5. Bot Detection*

Intrusion Detection System is also capable of differentiating the normal user requests from the requests generated by some software. A script can be created to either automate human activities on the web such as mouse clicks, keyboard strokes, etc. or launch web attacks including injection attacks, Dos attacks and brute force attacks. However, IDS should also have the potential to distinguish between good bots and bad bots. Good bots are those automated programs which are beneficial for the service provider, such as search engine bots which help in improving the ranking of a website.

In this section, we briefly discussed the security features provided by the IDS to prevent malicious activities on a website. Table 2 presents the list of web vulnerabilities which are covered by the different security measures.

**Table 2** Web vulnerabilities and security measures

| Security Measures | Web Vulnerabilities |
|---|---|
| Input Validation | Injection, cross-site scripting and unvalidated redirects and forwards |
| Output Validation | Injection, sensitive data exposure and using components with known vulnerabilities |
| Access Control | Broken authentication and session management, insecure direct object reference, security misconfiguration and missing function level access control |
| Session Verification | Broken authentication and session management and cross-site request forgery |
| Bot Detection | Insufficient monitoring mechanism |

## 5. IDS as Web Application Defence Mechanism

The section presents a systematic literature survey to provide a comprehensive overview of the current state of the intrusion detection systems in web application security. We study how the major approaches to intrusion detection have been utilized to deal with the complexities and issues of web applications, and what measures have been suggested to encounter the pitfalls associated with the detection approaches. Moreover, based on the review, a number of dimensions have been suggested to compare the existing systems and illuminate several pertinent facts.

*5.1. Signature-based Detection Systems*

As discussed in section 2.1, Signature-based systems detect intrusions by using the knowledge of existing attacks. However, the data being monitored can vary. For instance, the SIDS presented in study [52] analyses log entries to recognize malicious activities on the web server. The authors of the paper [48] showcase WebSTAT, an intrusion detection system based on STAT framework [11] which can monitor both logs and HTTP requests. The system relies on state transition analysis model to describe the attack scenarios and is also capable of detecting multi-step attacks. Moreover, besides being a misuse-based system, it permits detection of attacks variant which is similar to the specified malicious behavior. The authors in study [56] address the issue of manually writing the attack patterns by automating the whole process. The proposed work primarily makes use of text mining techniques to learn the behavior of both benign and malicious user from the web log file and generate the attack model.

---
[4] https://www.owasp.org/index.php/Top_10_2013-A4-Insecure_Direct_Object_References

*5.2. Anomaly-based Detection Systems*

The earliest work we came across in the literature that presented anomaly-based system which exclusively focuses on detecting web-based attacks is from Kruegel et al. [47]. In this paper, the authors have built multiple anomaly models to characterize the relationship between server-side programs, client-side parameters, and the corresponding parameter values. Several studies including [42], [51] and [53] used their work as the base for their research. In a study [53], authors introduce further features to capture more granular request behavior such as the number of times a program is invoked by the client (access frequency), if the requests are issued by an automated source (inter-request time) or if the client has bypassed any functionality (invocation order). In study [62], authors have presented the classification learning scheme to identify malicious behavior. In the classification scheme, the data of both the malicious as well as benign data is gathered to train the classifier. Research in a paper [65] has shown how DFA (Deterministic Finite Automata) can be used as an anomaly-based strategy to learn normal request behavior. Since user requests are highly variable in content, many transformation rules have been applied to these requests to reduce their variability. The transformation process reduced the complexity of the DFA model. Another technique, namely n-gram [7] is also seen as a promising technique to detect anomalies since it can analyze the structure of requests. However, the complexity of the n-gram modeling technique rises exponentially with the increasing size of the gram. The authors in study [64] proposed Spectrogram model that incorporates Markovian property to reduce the complexity of n-gram-based technique. Markovian property assumes the probability of future states depends solely on the present state. In one more study [60], the authors have used a group of Hidden Markov Model for detecting attacks related to the input validation. Authors in this paper have explicitly handled the problem of noisy data in the training set, which was earlier not focused by the researchers. The article [63] conducted a comparative analysis of two most promiscuous anomaly-based detection algorithms, DFA and N-grams based on their effectiveness and efficiency.

Some researchers have also attempted to introduce the context knowledge into the anomaly-based detector. For example, the study in paper [57] adds the notion of a context-aware anomaly detection system by incorporating the structural information of HTTP protocol into the technique. One-class support vector machine (OC-SVM) [16] has been used to build the anomaly detection model. The study [43] further carried the context-aware anomaly detection approach by presenting an innovative method for representing the packet payload using contextual n-grams ($_c$n-grams). The $_c$n-grams technique allows integration of structural properties of protocol and their respective byte sequences in a unified feature space. Furthermore, the concept of feature shading [2] has been presented to visualize anomalies which help to localize suspicious pattern and recognize the cause of an anomaly. Authors in work [51] have addressed one of the limitations of the anomaly-based approach, i.e., being inefficient in providing attack description.

Anomaly-based models are known for producing high false positive alerts, which took the attention of several researchers. Authors in paper [10] have proposed the idea of data compartmentalization to mitigate the problem of false alerts caused due to the requests which are anomalous but benign. Instead of following "all or nothing" methodology in which every anomalous request is treated as malicious, the requests are categorized on the basis of their anomaly score. Clients are given limited access (or no access) to the resources as per the category assigned to the requests. For example, the highly unusual request will be forwarded to the server that has no privilege to access the sensitive information. The same work is further extended in research [54] by serially integrating the web-based anomaly system with a database anomaly component. An IDS named Double Guard is proposed in paper [13] that considers both user requests arriving at the web server and database requests at the back-end to track the malicious requests. The system built the model by mapping client requests to their respective database queries. TokDoc [58], one more IDS is incorporated with automatic malicious request repairing mechanism. This innovative idea provides the ability to replace the suspicious section of the request with a safe value learned during the training phase. The TokDok is equipped with four healing strategies to deal with anomalous values: drop the token, encode the doubtful values, replace it with the most frequent values and change it to its nearest neighbor value. SWADDLER is another anomaly-based detection suggested by the authors in the study [50] which observes the internal state of the application by profiling the server side program variables. The approach is more granular as compared to the rest because it analyzes the different states of an application within a session while the attack is being performed. A similar approach, namely RRABIDS (Ruby on Rails Anomaly-Based Detection System Intrusion) [49] uses invariant logic to monitor the internal state of the application and recognize anomalies. The approach identifies a set of constraints on the internal variables which are dependent on the input values provided by clients.

There are several types of research in the literature which have exclusively focused on finding the solution to the issue arises in AIDS due to frequent changes in a web application. The work presented in paper [42] solves the problem of the higher number of false alerts raised when the application gets modified. The authors developed an innovative technique to automatically recognize the legitimate changes and selectively re-train the specific portion of the anomaly model. The work [44] is also dedicated to deal with the changes made in an application. To tackle the changes made in the application, the author applies transductive on-line learning property to build the model which is independent of the training set and improves over time.

### 5.3. Policy-based Intrusion Detection Systems

Researchers primarily use policy-based techniques because of their ability to set an equilibrium between allowed and not allowed events. Authors in the study [55] integrated a policy based Generic Authorization and Access control API (GAA-API) framework into an IDS to make the system more capable of identifying an unauthorized operation. The framework implements an Extended Access Control List (EACL) language [19] to specify security policies for monitoring resource access and performing actions in response to any threatening activity. Authors in the study [18] have focused on the ontological model as a new breed of IDS and highlighted two ontology models, namely protocol-centric and attack centric model to identify malicious requests. Protocol ontology model provides the baseline on which the security measures of the attack model are constructed. The model assists the system with inference and reasoning ability to map different scenarios of security breaches to a general semantic rule. It reduces the number of signatures in detection process. The authors in paper [14] further extends the work by using ontology engineering practices to design the system. Another study [17] embeds the Bayesian filter in an ontology-based system to mitigate the web attacks. The model includes both benign and malicious data in training to enhance the detection ability of the system.

### 5.4. Hybrid Intrusion Detection Systems

This section highlights the work on hybrid-based web IDS by different researchers so far. An intelligent Intrusion Detection and Prevention System (IDPS) proposed in paper [45] combines the anomaly-based and signature-based detection approaches along with additional response action mechanism to handle the intruders. The authors incorporate the DREAD model to estimate the threat risks and design the response policies according to the severity level. The IDS presented in [46] provides an architecture that leverages the strengths of both techniques (anomaly and signature) in such a manner that it gives the advantage of categorizing the events into safe, intrusive or unknown class (i.e., the class for which events neither qualify as an attack, nor as safe). Another hybrid detection approach presented in work [59] utilizes the features of the attacks performed by script kiddies. The misuse-based component of the system uses attack patterns provided by several web applications which participate collaboratively in the detection process. The web applications create a list of requests that they categorized as harmful and later on forward it to other cooperating web applications to enhance the detection process. The proposed system uses the weighted graph to recognize the anomalies.

Table 3 summarizes the contributions of various researchers towards intrusion detection systems designed particularly for securing web applications.

**Table 3** Summary of existing intrusion detection systems designed to secure web applications

| IDS | Contributions |
| --- | --- |
| Almgren et al., 2000 [52] | Presented a module-based SIDS. |
| Kruegel and Giovanni, 2003 [47] | Proposed AIDS exclusively for monitoring web application. Used statistical modeling techniques to profile features of request parameters. |
| Vigna et al., 2003 [48] | Proposed WebSTAT- an SIDS which is based on the STAT framework. Specified attack scenarios in terms of states and transitions. |
| Ryotov et al., 2003 [55] | Integrated Access Control module to IDS Proposed Generic Authorization and Access-control API (GAA-API) to identify an unauthorized operation |
| Tombini et al., 2004 [46] | Proposed HIDS by combining anomaly and misuse detection approach. Categorized the requests into 3 classes, namely safe, intrusive and unknown class. |
| Kruegel et al., 2005 [53] | Extended the work presented in [47] by providing three additional features to profile the relationship between requests. |
| Robertson et al., 2006) [51] | Overcame the limitation of AIDS approaches of insufficiency in providing attack description<br>Used generalization technique and regular expressions to group anomalies and infer attack classes respectively. |
| Valeur et al., 2006 [71] | Proposed the concept of data compartmentalization to reduce false positives in AIDS.<br>User requests are handled based on their anomaly scores. |
| Adeva and Atxa, 2007) [56] | Proposed text mining based SIDS.<br>Automated the process of signature creation using text categorization. |
| Cova et al., 2007 [50] | Proposed characterization of the internal state of the web application to detect anomalies.<br>Used both multivariate and univariate models for profiling. |
| Ingham et. al., 2007 [65] | DFA modeling has been used in anomaly detection method<br>Applied several heuristics measures to anomalous requests to reduce false positives. |
| Dussel et al., 2008 [57] | Incorporated the concept of HTTP protocol into anomaly detection technique.<br>Adopted one-class support vector machine (OC-SVM) to build the AIDS. |
| Park et al., 2008 [12] | Used Needleman-Wunsch algorithm [15] from bioinformatics to build the AIDS. |
| Maggi et al., 2009 [42] | Proposed AIDS that deals with continuous changes in a web application.<br>Used HTTP Response data to detect changes. |
| Vigna et al., 2009 [54] | Integrated web-based and database-based anomaly detection systems |
| Song et al., 2009 [64] | Proposed factorized n-gram Markov technique based AIDS that reduces the complexity in n-gram method. |
| Corona et al.,2009 [60] | Used Hidden Markov technique to build AIDS.<br>Explicitly dealt with noise in training data. |
| Razzaq et al., 2009 [17] | The Bayesian filter is applied to ontology system to mitigate web application attacks. |
| Kruegel et al., 2010 [58] | Presented several healing strategies to recover malicious requests.<br>Replaced suspicious section in a request with benign data on the basis of previously trained anomaly detectors. |
| Corona et al., 2010 [61] | Extended the work proposed in the study [60] by using statistical models along with HMM model to enhance detection efficiency.<br>Used clustering technique to identify anomalies. |
| Lin et al., 2010 [63] | Performed comparative analysis on two promiscuous anomaly-based detection algorithms, namely DFA and N-grams in terms of effectiveness and efficiency |
| Lampesberger et al., 2011 [44] | Addressed the tendency of continuous changes in web application<br>Introduced transductive on-line strategy to train AIDS. |
| Ludinard et al., 2012 [49] | Proposed invariant based anomaly detection model.<br>Focused on monitoring of violations of the internal state of application |
| Lee et al., 2012 [13] | Used container-based architecture in AIDS.<br>Mapped HTTP requests to database queries. |
| Wressnegger et al., 2013 [62] | Performed comparative analysis on two learning schemes, namely classification |
| | and anomaly detection. <br> Also defined criteria for selecting the appropriate scheme. |
| Alazab et al., 2014 [45] | Built IIDPS by embedding SIDS with AIDS. <br> Proposed active response strategy using fuzzy logic <br> Used DREAD model to assess the risk associated with alerts. |
| Razzaq et al., 2014 [18] | Proposed semantic-based approach that uses ontologies to detect web-based attacks <br> Provided protocol-based and attack-based ontology model |
| Razzaq et al., 2014 [14] | Suggested how ontology-engineering practices could be applied to design ontology-based systems. |
| Duessel et al., 2016 [43] | Modelled context-aware anomaly detection system by using $_c$n-gram method. |
| Marek Zachara, 2016 [59] | Proposed HIDS and used weighted-graph as an anomaly method. <br> Misuse detection component uses attack patterns provided by different websites contributing to the detection process. |

## 6. Dimensions of web IDS

The IDS tools are generally compared on the basis of performance metrics such as detection coverage, false positive rates, etc. [94]. Here, we are interested in comparing the detection systems which are designed exclusively for preventing web-based attacks on the basis of their design methodology, and the security features and functionalities they offer. After critically and comprehensively reviewing the literature, several dimensions have been recognized so as to compare the existing web IDS from multiple perspectives. Each dimension comprises a number of attributes which provide a firm baseline for the comparison. The dimensions are briefly described as under:

### 6.1. Detection Approach

IDS can be categorized on the basis of detection technique that has been employed to recognize security breaches. The detection techniques include anomaly, signature, policy and hybrid approach as discussed earlier in section 2.

### 6.2. IDS Type

IDS can also be categorized into HIDS and NIDS on the basis of its operating layer (application or internet) of TCP/IP. HIDS monitors the data flow (in/out) of the terminal upon which IDS has been configured. For the web applications, HIDS mostly operates as a reverse proxy and are usually placed on the web server that hosts the website. On the contrary, NIDS is deployed at strategic points in the network infrastructure so that it can see the data flowing in the network. Since web attacks mostly target application layer and NIDS lacks the information regarding application context, these systems are generally not considered as an excellent choice for detecting web application attacks. Furthermore, as already discussed in section 3.1, they are not capable of efficiently monitoring the encrypted traffic.

### 6.3. Data Monitored Type

Intrusion detection systems can also be classified according to the kind of data they are inspecting. Several types of data being monitored by different IDS are mentioned as under:

*6.3.1 Logs:* Most of the detection systems process by analyzing the log files generated by the web server [21]. Logs are usually maintained to keep the record of the client request. However, these files do not contain full information of the data that is sent to the server from the client such as data passed in HTTP header and POST request.

*6.3.2 HTTP Header:* IDS observes the metadata of the request passed in an HTTP header.

*6.3.3 GET Request Data:* Detection system analyzes those requests which use GET method to send the data to the server. In other words, IDS examines the query string contained in the URL.

*6.3.4 POST Request Data:* The system analyzes those requests which use POST method to send the data. In the POST method, client data is embedded in the message body of a request. These requests allow the users to pass a more substantial amount of data to the server.

*6.3.5 Application variables:* Web applications consist of a number of server-side programs to process the requests, and the application variables are the variables used in these programs. Monitoring these variables make detection approach more granular and effective.

*6.3.6 Response Body:* It consists of content sent back by the server to the client. It assists the IDS to monitor the flow of data from the server to the client, unlike the other data types which are used to control the data flowing from client to server.

*6.4. State*

IDS can either be stateful or stateless. As already discussed in section 3.3, the stateless IDS analyzes incoming requests independently, whereas the stateful IDS can efficiently understand the relationship between multiple requests. The stateful analysis is more powerful and capable of detecting attacks since it can observe the activities within a session. It also assists in recognizing the attacks which use a series of requests to exploit the vulnerability.

*6.5. IDS Mode*

IDS are supposed to monitor web applications and identify any sign of incidents which try to misuse the applications. Upon encountering any abnormal or misleading event, the system usually logs the event or alerts the administrator. These systems lack the prevention mechanism. There also exists an enhanced version of the system called Intrusion Prevention Mechanism (IPS) which can identify as well as prevent the adversarial activities such as rejecting the current client request, blocking the client, etc. [35]. Here, we compare the IDS based on whether they also include prevention mechanism in addition to default detection capability. The ability to respond to malicious actions makes the prevention systems more desirable than the ones which only detect

*6.6. Response Mechanism*

Response mechanism is the process of taking actions against intrusive activities. It can be either passive or active [22]. In passive response, the IDS does not take any corrective step themselves to lessen the damage caused by the attacks. Their mere job is to log the security breach and notify the administrator, whereas in the active-response mechanism, the IDS can take corrective action in response to the security compromise. The active responses may include blocking the attacker's IP address, refraining attackers from performing further activities or resuming the service of suspended accounts. The IDS with no preventive measures responds passively, whereas the system with the ability to take countermeasures reacts actively. Furthermore, active responses can be selected on the basis of either static mapping or dynamic mapping [66]. In the case of static mapping, the response is independent of the context of attack and action is triggered based on the predefined rigid response-rules. On the other hand, the responses based on dynamic mapping take attack context into consideration and action is triggered based on the evaluation of several attack metrics such as severity level, risk, etc.

*6.7. Attack Description*

This dimension groups the IDS which provide a description of the attacks on alerts. Characterizing the attacks helps the administrator to visualize the intrusion and identify the vulnerabilities existing in the applications. The information obtained from description may also be leveraged to prevent attacks in future.

*6.8. Incremental Learning*

Learning in complex and changing environment demands for those systems which can update themselves with the change in the application. Hence, the detection systems can also be compared based on whether they can update themselves with time, and are impervious to modifications to applications.

*6.9. Security Measures*

As already discussed in section 4, the security measures offered by existing IDS include input validation, output validation, access control, session verification and bot detection. All the detection systems usually do not incorporate all the mentioned security measures. So, we can also differentiate the IDS in terms of the kind of security features they are providing.

# 7. Comparison of IDS

In this section, we provide a comparative study of the existing web IDS mentioned in the literature from identified dimensions. Each IDS has been assessed by a respective set of features of the individual dimension. Table 4 shows the comparison of existing systems.

**Table 4** Comparison of IDS

| Reference | Detection Approach | IDS Type | Data Monitored Type | State | IDS Mode | Response Mechanism | Attack Description | Incremental Learning | Security Measures |
|---|---|---|---|---|---|---|---|---|---|
| [52] | SD | HIDS | Logs | Stateful | Detection | Passive | No | No | IV AC BD |
| [47] | AD | HIDS | Logs, Get | Stateless | Detection | Passive | No | No | IV |
| [48] | SD | HIDS | Logs, Get, Post, Header | Stateful | Detection | Passive | No | No | IV AC SV |
| [55] | PD | HIDS | Get, Post, Header | Stateful | Prevention | Active & Dynamic | Yes | No | IV AC |
| [46] | HD | HIDS | Get, Header | Stateless | Detection | Passive | No | No | IV |
| [53] | AD | HIDS | Logs, Get | Stateful | Detection | Passive | No | No | IV BD |
| [56] | SD | HIDS | Logs | Stateful | Detection | Passive | No | No | AC |
| [51] | AD | HIDS | Logs, Get | Stateless | Detection | Passive | Yes | No | IV |
| [71] | AD | HIDS | Get | Stateless | Prevention | Active & Dynamic | No | No | IV |
| [50] | AD | HIDS | Application Variables | Stateful | Detection | Passive | No | No | IV AC |
| [65] | AD | HIDS | Get, Post, Header | Stateless | Detection | Passive | No | Yes | IV |
| [57] | AD | NIDS | Get, Header | Stateless | Detection | Passive | No | No | IV |
| [12] | AD | HIDS | Get, Post, Header | Stateless | Detection | Passive | No | No | IV |
| [42] | AD | HIDS | Get, Post, | Stateless | Detection | Passive | No | Yes | IV |

| Ref | | | | | | | | | |
|---|---|---|---|---|---|---|---|---|---|
| | | | Response | | | | | | |
| [54] | AD | HIDS | Get | Stateless | Prevention | Active & Dynamic | No | Yes | IV AC |
| [64] | AD | NIDS | Logs, Get, Post | Stateless | Detection | Passive | No | No | IV |
| [60] | AD | HIDS | Get | Stateless | Detection | Passive | No | No | IV |
| [17] | PD | HIDS | Get, Post, Header | Stateless | Detection | Passive | No | Yes | IV |
| [58] | AD | HIDS | Get, Post, Header | Stateless | Prevention | Active & Dynamic | No | No | IV |
| [61] | AD | HIDS | Get, Header | Stateless | Detection | Passive | No | No | IV |
| [63] | AD | HIDS | Get, Post, Header | Stateless | Detection | Passive | No | No | IV |
| [44] | AD | NIDS | Get, Post, Header | Stateless | Detection | Passive | Yes | Yes | IV |
| [49] | AD | HIDS | Application Variables | Stateful | Detection | Passive | No | No | IV AC |
| [13] | AD | HIDS | Get, Post, Header | Stateful | Detection | Passive | No | No | AC SV |
| [62] | AD | NIDS | Get, Post, Header | Stateless | Detection | Passive | No | No | IV |
| [45] | HD | HIDS | Get, Post, Header | Stateless | Prevention | Active & Dynamic | No | No | IV BD |
| [18] | PD | HIDS | Get, Post, Header, Response | Stateful | Prevention | Active & Static | Yes | No | IV AC SV OV |
| [14] | PD | HIDS | Get, Post, Header | Stateless | Prevention | Active & Static | Yes | No | IV |
| [43] | AD | NIDS | Get, Post, Header | Stateless | Detection | Passive | Yes | No | IV |
| [59] | HD | HIDS | Get, Post, Header | Stateful | Detection | Passive | No | Yes | IV BD |

AD: Anomaly-based Detection  
SD: Signature-based Detection  
PD: Policy-based Detection  
HD: Hybrid-based Detection  
HIDS: Host-based Intrusion Detection System  
NIDS: Network-based Intrusion Detection System  

IV: Input Validation  
OV: Output Validation  
AC: Access Control  
SV: Session Verification  
BD: Bot Detection  

Based on the comparison, we evaluate the contribution of each feature to their respective dimension. The state of each dimension is illustrated in Figure 1 that illuminates several pertinent facts about the current state of the literature. As clearly visible from the *detection approach* dimension, the representative sample of works (67% of the total) has focussed on anomaly-based IDS. Anomaly-based approach assists in developing intelligent IDS with the ability to detect novel attacks as well. However, the current research is attracted towards framing either hybrid or policy-based systems. Policy-based systems impose a set of rules to establish a balance between the anomaly and misuse detection techniques. Likewise, hybrid-based systems are capable of clubbing the strengths of multiple detection techniques in a way to overcome the limitations of one approach by another. Another dimension, namely *IDS type* shows that most of the detection systems are host-based (83%) as compared to network-based systems (17%). It can be realized from the fact that NIDS is not capable of inspecting HTTPS traffic until it has SSL certificate key. NIDS also lacks the context of web application technology. When it comes to *security measures* dimension, the input validation is the leading feature that almost every detection system is providing (93%). Inadequate input validation is the prime reason behind most of the applications to be vulnerable.

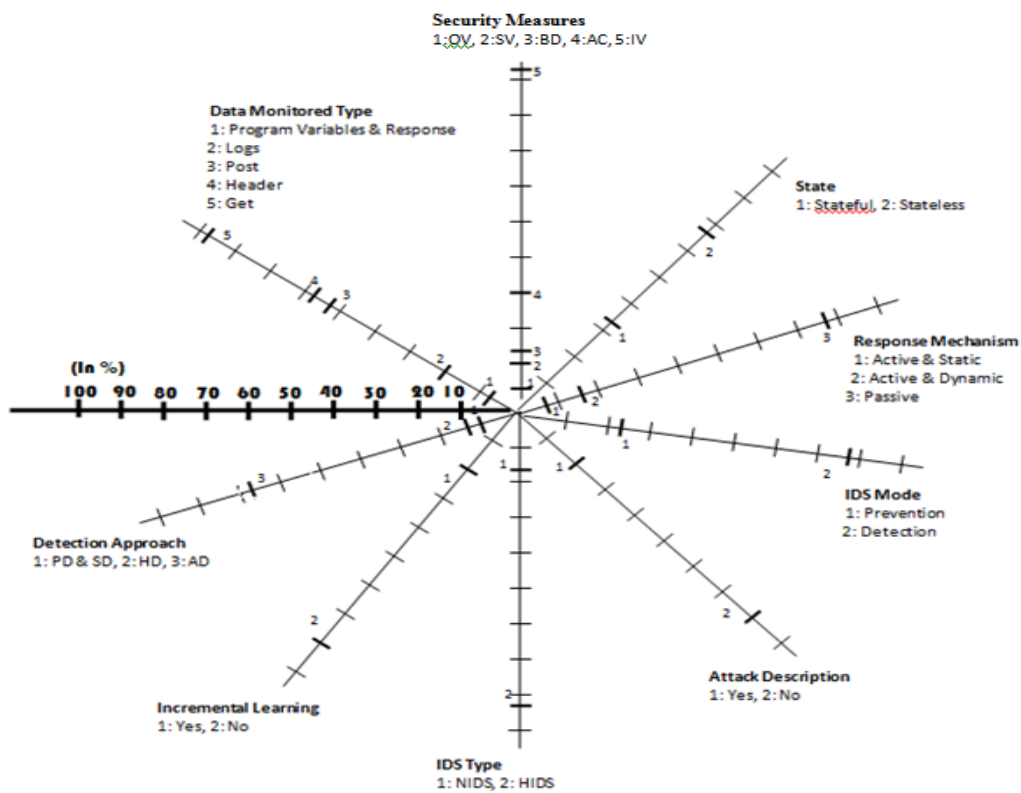

*Fig.1.* Intrusion Detection System Dimensions

However, the graph also enlightens the features of several dimensions which need to be highly explored by the researchers. For instance, *data monitored type* dimension shows, only 3% of the systems are monitoring the traffic flowing out of the web server. The effect is also reflected in the security measures dimension where 3% IDS offer output validation. As we discussed in section 4.2, monitoring the response data refrain the application from exposing sensitive information unintentionally. Moreover, the parameter tampering based attacks can only be checked if the IDS watches outflow of data. The *security measures* dimension also reveals that only 13% of systems have been designed to monitor bot activities. According to a statistical report on the bot traffic[5], every third visitor on the website is an attack-bot which indicates how alarming it is for the IDS to concentrate on detecting requests from the automated programs. It is also shown in the graph via *state* dimension that the existing systems have overlooked the tracing of the relationship between different requests, and are more stateless (67%) than stateful (33%). The stateless detection system cannot track the behavior of a particular client, and therefore, fails them to recognize certain critical malicious behaviors such as session-based attacks and unauthorized access to resources. Session hijacking is one of the vulnerabilities leveraged prominently by the hackers to conduct fraud online money transactions[6]. According to the 2017 WhiteHat Security report, insufficient authorization is one of the top classes of vulnerabilities found as critical. Since stateless systems are not informed of session Ids, they cannot also counteract vigorously against the adversaries such as blocking/suspending the offending session or user. They operate at request level, and therefore, can only deny the request at maximum as a preventive strategy.

The graph further illustrates that the prevention mechanism in the IDS has received less attention. The majority of the systems are passive (77%), therefore lacks the functionality to take appropriate measures in response to a threat. These systems are only concerned with recognizing and reporting the malicious activities. However, identifying an intrusion is just half the battle, knowing how to respond is equally essential. Once a suspicious activity has been detected, it is desirable to have an automated resource capable of taking appropriate steps to mitigate the damage caused by the attack [96]. The *alert description* dimension shows that the researchers also need to pay attention to the IDS which provides the context of the generated alerts. Only 20% systems have been seen with such a feature. Investigation of the alarms raised by the system is an essential task. It helps the administrator to get a condensed view of the security issues which assists in studying the cause of the alert, reducing the false positive rate, and adopting suitable countermeasures. Therefore, the management of alerts is one of the prevalent research fields in the intrusion detection domain [97]. However, for the tasks to be productive, a good deal of information regarding alert is needed. It reduces the effort spent by the analyst on delivering timely and effective countermeasures [98]. Lack of proper description of alarms makes the task time consuming and resource-intensive. Another dimension which lacks the attention of researcher is *incremental learning*. The systems which adapt to changes are proved to be intelligent enough to learn and operate in the complex and dynamic environment. Only one in five IDS is equipped with the dynamic learning capability. As we discussed in section 3.4, web applications undergo several changes over time, and therefore it is not necessary for the IDS to perform as good in future as it is working at present. Automatic learning capability ensures that the profile is up-to-date.

---

[5] https://www.incapsula.com/blog/bot-traffic-report-2016.html
[6] https://cdn2.hubspot.net/hubfs/2264844/website/pdf/anatomy_of_banking_fraud.pdf?t=1509722403738

The fundamental aim of showing this graph is to present the current state of the art in intrusion detection systems for web applications. It provides a comparative glance at different existing IDS based on several dimensions. In a nutshell, the graph highlighted the areas where the research efforts are still isolated. It would highly facilitate the researchers and academicians to understand the stronger as well as weaker aspects of the domain.

## 8. A conceptual framework of IPS for web applications

Although intrusion detection is a well-known methodology in preventing adversarial activities on the network, it is still naive in the domain of safeguarding web applications. In section 3, we discussed several challenges which researchers face while building the web IDS, and in section 6, we talked several dimensions for comparing the existing IDS. In this section, we provide a conceptual framework of an intrusion detection system along with prevention mechanism that caters all the challenges being discussed while reviewing the literature. The proposed framework offers the systematic guidance for the implementation of the system exclusively for web applications. The AppSensor project of OWASP [89] also presents a framework that assists in implementing the IDS for web applications. Thus, to showcase the novelty of the proposed framework, we compare its features with the AppSensor framework and one another intrusion detection tool, PHPIDS [90]. Since today's modern web application firewalls are offering similar security services, the proposed system is also compared with three commonly used open source WAFs, namely ModSecurity [91], Shadow Daemon [92] and AQTRONIX WebKnight [93]. The proposed IPS acts as a reverse proxy that intercepts both incoming requests and outgoing responses from the client and server respectively. Following are given the essential design features of the proposed framework:

- The framework of proposed IPS adopts hybrid detection approach for utilizing the capabilities of both signatures and anomaly-based detection techniques. The signature detection methodology helps to define stringent rules for both whitelisting and blacklisting the already known content, and anomaly detection methodology facilitates the detection system in learning the normal application behavior.
- The presented framework follows modular architecture where the whole system is divided into five components, namely *Preprocessor, Detector, Defender, Logger* and *Response Controller*. Each component is further divided into its respective modules and sub-modules.
- The IPS stores the configuration and behavioural profiles of each web application as per its business logic to understand its structure, functionalities, and operations. The configuration steps and building of behavioural profiles are explained along with the components of the framework.
- The IPS also includes the SSL offloading feature to encrypt and decrypt the SSL traffic. It receives both HTTP and HTTPS requests, decrypts the content if the request is HTTPS, verify the content and forwards non-malicious requests to the server in HTTP format. Similarly, the IPS receives the HTTP responses from the web server, processes the content, encrypts the content for HTTPS communication and finally sends it to the client. The *preprocessor* component handles the decryption, whereas *response controller* component provides the encryption functionality.

Figure 2 depicts the block diagram of the working of the IPS. Table 5 compares the design features of the proposed one with the detection tools mentioned above. The working of each component is discussed in the following sub-sections.

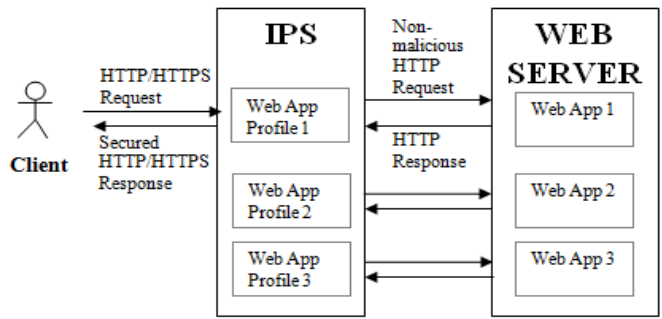

*Fig.2.* Block Diagram of the IPS working

Table 5 Comparison of frameworks in terms of their design features

| Feature | AppSensor | PHPIDS | ModSecurity | Shadow Daemon | Aqtronix WebKnight | Proposed IPS |
|---|---|---|---|---|---|---|
| Detection Methodology | Hybrid | Signature (Blacklist) | Policy | Signature (Blacklist and Whitelist) | Signature (Blacklist) | Hybrid |
| Modular Design | Yes | No | No | Yes | No | Yes |
| Application Specific | Yes | No | Yes | Yes | Yes | Yes |
| SSL offloading | Yes | No | Yes | Yes | Yes | Yes |

*8.1. Preprocessor*

A preprocessor is usually the first component of every IDS which is responsible for converting the input data into some standard format. However, the proposed preprocessor includes one more functionality whose job is to pass only necessary requests to the IDS. In particular to web applications, the HTML response usually contains a number of references to the static objects such as images, CSS files, and JavaScript files. When the HTML page is rendered by the web browser, it makes separate HTTP request for every single file referred in the page. The IDS should be programmed to ignore such requests since the detector has no significant role in monitoring these requests. The practice will significantly reduce the load on IPS and enhance the performance of the website. Therefore, the proposed pre-processor component consists of two modules viz *Requests Filter* and *Data Constructor*, which are described as under.

*8.1.1 Request Filter:* The job of the filter is to prevent those requests from the monitoring mechanism which generated while rendering a web page such as image requests, script file requests, etc. These requests can be passed directly to the web application for processing. If every request goes through the detector, it may cause unnecessary delay in the loading of a web page on the browser. The Request Filter needs to be trained to learn the requests to be filtered. During the training phase, when the application is being accessed under safe environment, the model can be built to learn about the requests sent from the client and the requests submitted on behalf of the application.

*8.1.2 Data Constructor:* The requests to be monitored are forwarded to the second module of the Preprocessor, the Data Constructor. It prepares the data in a format that will be used by the Detector component for analyzing the requests. It reads the raw content of HTTP request and structures it into several fields to be investigated for recognizing a suspicious request such as source IP address, the timestamp of the request, HTTP method, URL requested, HTTP headers, cookies, post request data, get request data, etc. The module also applies various transformation functions to reduce the variability in the text content such as removing extra white spaces, converting the content into the same case (either upper or lower case), decoding the data, etc. The transformation functions help the detector in two ways. 1) Enhance its performance by supplying an optimal data, and 2) assist in dealing with the IDS evasion techniques like encoding schemes (UTF-8, hex or Unicode[7], etc.). The sub-module is also assigned the task to decrypt the data sent via SSL to inspect the HTTPS request as well.

In Table 6, we compare the pre-processor component with the other five systems. It can be seen that none of them is offering the functionalities of Request Filter module. Instead, they monitor every request which may cause a delay in the loading of a web page if the number of static object requests would be high in an HTML page.

---

[7] http://www.cgisecurity.com/lib/URLEmbeddedAttacks.html

**Table 6** Comparison of *Preprocessor* component feature with others

| Module | AppSensor | PHPIDS | ModSecurity | Shadow Daemon | Aqtronix WebKnight | Proposed IDS |
|---|---|---|---|---|---|---|
| Request Filter | No | No | No | No | No | Yes |
| Data Constructor | Yes | Yes | Yes | Yes | Yes | Yes |

*8.2. Detector*

The Detector is the most crucial component of the IDS as it is primarily responsible for determining intrusive requests. The job of the Detector is to analyze the HTTP requests from several aspects mainly regarding its content and behavior. The functionality of the Detector is divided into five modules, namely Connection Verifier, Bot Detector, Data Validator, User Verifier and Access Controller. The modules are serially connected to each other in the specified sequence. If any of the modules found a suspicious request, it forwards the request to the Defender component (the component that handles the malicious requests), or else to the subsequent module of the Detector for further verification. The working of each module is discussed below.

*8.2.1 Connection verifier:* It is the first module in the Detector component whose job is to recognize those requests which are either originated from the source or destined to the target that has been either blocked or suspended for a particular period. The module is implemented via a blacklist of the following objects.

- The list has a record of sources which are either blocked or suspended for some time being. There are certain cases when an application denies the requests from a particular client. For example, once the detector determines the requests are coming from a computer program (bot), the IPS may decide to reject further requests from that source. It helps in protecting the website from already identified adversaries, site scrapers, or spammers by blocking them or suspending them for a specific duration. The blacklist comprises a number of identifying details of the source such as the username of the account holder, the session id of the client, IP address, etc.
- The module also contains a list of services (targets) which are made temporarily/permanent unavailable due to some technical or security reasons. There are specific scenarios when a web application wants to restrict the access to a particular service, for example, when the site is under maintenance [75].
- Besides, the module also contains the list of files and directories to be prevented from direct access. For instance, there can be some sensitive files (such as config file which includes the entire system configuration) or directories (such as "includes" directory which generally contains the scripts used in the web pages) for whom the application wants to restrict direct access over the web.

The blacklists of the Connection Verifier module can be maintained either manually by an administrator or by the Defender component of the IPS. For instance, once the Defender component decides to block or suspend a specific user session, an IP address or a service, it may update the Connection Verifier module to detect subsequent requests for the same. If the request passes the Connection Verifier, it is sent to the next module, Bot Detector for further verification, or else is sent to the Defender Component to deal with the suspicious request. Table 7 compares the proposed IPS with other five systems from the features provided in Connection Verifier.

**Table 7** Comparison of *Connection Verifier* module features with others

| Security Features | AppSensor | PHPIDS | ModSecurity | Shadow Daemon | Aqtronix WebKnight | Proposed IPS |
|---|---|---|---|---|---|---|
| Blacklist blocked /suspended source | Yes | No | Yes | No | Yes | Yes |
| Blacklist blocked /suspended target | No | Yes | Yes | Yes | Yes | Yes |

*8.2.2 Bot Detector:* As already discussed, the HTTP requests may also be issued by some software programs. The main task of this module is to determine whether the requests have been issued from a human or an automated program. It is to be noted that not all bots are bad. There are certain good bots such as GoogleBot [76] which are necessary for increasing the ranking of a website in search engines. In order to detect malicious bots, the module needs to understand how to differentiate between humans, good bots and bad bots. Therefore, the Bot Detector module comprises three sub-modules. In the first sub-module, the requests sent by non-malicious bots are filtered based on their IP addresses that specify the address of the web hosting services used to generate bot traffic. There are websites which provide the address of the bots to be whitelisted[8]. It is mainly the job of the administrator to form the list of IP addresses of the permitted bots. The request sent from non-malicious bots are forwarded to next module of the Detector, the Data Validator for further verification of suspiciousness, otherwise, passed to the subsequent sub-module of Bot Detector.

In the second sub-module, the IP address along with various header details such as user-agent are examined to identify the requests from bad bots. User-agent contains information regarding the browser and the technology used to access the website. There are several websites which publish the user-agent and IP addresses of the known malicious web robots [72, 73]. If the request passes through the second sub-module, it enters the third sub-module where the behavioral analysis of the request is carried out.

Technical details of a request are not enough to distinguish a bot activity from the human. One of the reasons is that bots can easily hide their identities by using spoofed values of the request elements and bypass the second sub-module. Hence, there is a third sub-module that monitors the behavioral patterns of the incoming requests to spot them. The sub-module is first trained to learn the normal usage of an application. It uses a combination of distinguishing characteristics to determine the likelihood of a request being sent from an automated agent. Various signs signify the use of bots on an application. Foremost is the request rate. Bots generally issue HTTP requests at a much faster rate than a human. By tracking the IP address of the client and analyzing the average speed with which it is sending requests, the module can detect a bot. Next parameter can be the number of requests sent from an IP address in a particular duration since bots can generate thousands of requests in a short period. The navigational pattern can also be used to serve the purpose since the website is accessed differently by bots. The number of hits that generate errors is another feature, as a bot looking for application flaws tend to generate a significant amount of 404 pages. Other characteristics used to identify the bot traffic include types of resources requested, the time interval between queries, coverage of the web application, etc. [69, 70, 71]. If the request is found malicious, it is sent to the Defender component to take appropriate actions. The Defender may decide to block or suspend the source to deny the further requests. The component will make the required changes in the blacklist of the Connection Verifier. The systematic diagram of the working of the Bot Detector module is shown below in Figure 3.

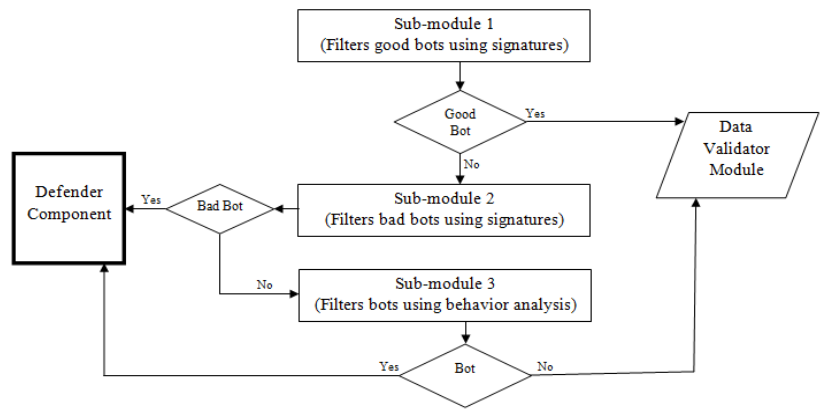

*Fig.3.* *Working of the Bot Detector module*

Table 8 shows the comparison of the security features provided by the Bot Detector with the other systems. The signature-based tools including PHPIDS, Shadow Daemon and Aqtronix WebKnight lack the functionality to monitor the behavior of requests. Besides, both PHPIDS and Shadow Daemon do not also track the IP addresses. In the case of the AppSensor framework, there is no security policy which whitelists the good bots. ModSecurity is the only tool that provides all the features but it requires the administrator to specify the malicious behavior of the bots in the form of rules, on the contrary, we proposed sub-module 3 that uses machine learning techniques to learn malicious behavior.

Table 8  Comparison of *Bot Detector* module features with others.

| Bot Detector Module | AppSensor | PHPIDS | ModSecurity | Shadow Daemon | Aqtronix WebKnight | Proposed IPS |
| --- | --- | --- | --- | --- | --- | --- |
| Sub-module 1 | No | No | Yes | No | No | Yes |
| Sub-module 2 | Yes | No | Yes | No | Yes | Yes |
| Sub-module 3 | Yes | No | Yes | No | No | Yes |

---

[8] https://www.ip2location.com/free/robot-whitelist

*8.2.3 Data Validator*: This module secures the websites from the attacks which exploit the vulnerabilities proliferated due to the improper implementation of input validation. The module validates the client data including the header, query string and form data. Before validating any value, the module first determines the type of the input parameter. From the characteristics of values, the module distributes the parameters into six categories, namely text, numeric, enumerated, format-specific, web address and application parameters. Based on the type of input parameters, the values are validated. The validation schemes of different parameters are discussed as follows:

- Since *text* parameters can contain an indefinite set of values, they are verified using the blacklist which includes the signatures of already known attacks such as SQL injection and XSS-based attacks in order to ensure that the value does not contain any potential malicious string. The blacklist is created manually and needs to be updated as the new attack pattern arrives on the web.
- For *numeric* category, the Validator module puts the request in the suspicious class if the parameter contains non-numeric values.
- The *enumerated* parameters assume values from a pre-defined set which are validated using white-list. Manually, it is not possible to create white-lists for all the input fields since each white-list will have a different set of values and the number of input fields can be many depending on the business logic of the application. However, machine learning techniques can be used to train the Validator to learn the set of values and generate the individual white-list. In paper [53], the author discovers the parameters that can have limited number of values by determining if the number of distinct values of the parameters is bounded by some threshold, and builds the set of legitimate values for such parameters.
- Similarly, for the *format-specific* category also, machine learning can be applied to learn the structure of the values and generate the pattern for verifying the parameters. For example, the authors in the paper [47] capture the structure of the parameter values by learning the distribution of relative character frequencies.
- The parameter which contains a *web address* (i.e., a URL), is primarily verified to prevent the application from unvalidated redirects and forwards vulnerability threats. Some web applications include code that redirects the user to another web page of the same domain or different domain based on the client side parameter. Here, the Validator needs to ensure that the parameters do not contain any malicious or unwanted link of a web page. The best possible way to prevent such attacks is to sanitize the input by using the whitelist that has the details of all trusted URLs [99]. The list of valid URLs can either be configured manually or generated via machine learning techniques by learning the valid URLs space from trusted traffic during the training phase.
- The last category, *application* is comprised of parameters whose values have been initialized by server-side programs and must not be modified while transmitted back from the client. The Validator module takes the help from Response Controller component of IPS to validate these values. The Response Controller marks the values which are set by application and store the hash of these values in a list. The Data Validator then uses this list to verify whether the values are altered or not. Table 9 contains the description of each of the parameter categories along with their validation scheme.

**Table 9** Description of parameter categories

| Parameter category | Description | Examples | Validating Scheme |
|---|---|---|---|
| Text | Holds both letters and numbers in the indeterminate range. | Username= John Smith Address= Celeste Slater 606-3727, Roseville NH11523 | Blacklist of prohibited content |
| Numeric | Holds only numbers. | Id=1254, Age=23 | Ensuring valid data type |
| Enumerated | Consist of values of a fixed set of elements defined by the application business logic. | Gender={'male','female','other'} Married={'yes','no'} | White-list of allowed values |
| Format-Specific | The values which follow a rigid pattern. | Date= Nov 4, 2003 8:14 PM Time= 8:14:11 PM | White-list of valid patterns |
| Website address | Holds URL as a value. | Target=www.partnersite.com Fwd=appadmin.jsp | White-list of valid URLs |
| Application | The values which are set by application and must not be tampered at the client side. | JSESSIONID=ABAD1D Price=12.3 | A list containing hashes of these values |

If any of the client-side values fails the Validator checks, the request is sent to the Defender component. Depending on the severity of the malicious request, the Defender may reject the request or pass the request to the application by just logging the event details. Table 10 represents the availability of these features in other detection systems. The tools, PHPIDS, ModSecurity and Aqtronix WebKnight are not concerned about the context of the parameters. Their architecture does not support the way through which different fitter rules can be designed for different types of parameters.

**Table 10** Comparison of *Data Validator* module features with others

| Validated Parameter Type | AppSensor | PHPIDS | ModSecurity | Shadow Daemon | Aqtronix WebKnight | Proposed IPS |
|---|---|---|---|---|---|---|
| Text | Yes | Yes | Yes | Yes | Yes | Yes |
| Numeric | No | No | No | Yes | No | Yes |
| Enumerated | Yes | No | No | No | No | Yes |
| Format-specific | No | No | No | Yes | No | Yes |
| Website Address | Yes | No | No | No | No | Yes |
| Application | Yes | No | No | No | No | Yes |

*8.2.4 User Verifier:* The attacker can also exploit the web flaws to impersonate a user and make requests on his behalf. User Verifier module monitors the activities of the user to ensure that the owner of the account is issuing requests. The module contains two sub-modules for detecting session impersonation attacks. The first sub-module examines various session related details such as the number of login attempts, IP address, user-agent, the idle period, anti-CSRF token value, etc. to prevent an attacker from accessing the web application by using another user identity. The user can be treated as suspicious if he successively fails in logging into his account. The use of different IP addresses and User-agents for the two requests of the same session may raise the doubt of the session being used by more than one user. Similarly, the session can be made expire after a suitable period of inactivity to tighten the security as it would give the attacker a smaller window of opportunity to get a valid session identifier. The anti-CSRF token is a random value which can be added by the Response Controller component of the IPS to the response body as a hidden field in the HTML form or as a query string parameter in the links. When a client issues a request, the sub-module examines whether it includes the embedded token and has the same value. Examining the anti-CSRF tokens would prevent the application from Cross-Site Request Forgery-based attacks [82]. If the request fails any of the checks specified in the first submodule, it is forwarded to Defender Component for further action, or else is sent to next sub-module of the User Verifier.

The mitigation approach used by the first sub-module to detection of session impersonation will significantly fail for two major reasons. 1) The victim may use public IP address which multiple machines may share including the attacker, and 2) The attacker may spoof IP address, user-agent and other details of the victim. Therefore, the second sub-module analyzes the user behavior and detects the attacks by inspecting the significant changes in the behavior of the session being operated. The module monitors the set of actions based on the premise that masquerader will perform operations that sufficiently deviate from the typical profile behavior of the victim. The module is first trained to distinguish legitimate user behavior from the malicious one. Different behavioral characteristics such as time of the session, kind of activities performed on the application, the frequency of the requests, request sequences, temporal dependencies among requests, etc. [77, 78, 79, 80] can be used for user profiling. If the request conforms to the established user profile, it is sent to next module of the detector, the Access Controller, otherwise to the Defender. Moreover, on a valid request, the module gets the new data which is used to update the corresponding user profile. In the case of session impersonation, the Defender may make the user immediately log out of the application and notify the account holder about the security breach. It may also suspend the account or in certain critical security applications (e.g., online banks) even make the account disabled until the user activates it through some out-of-band-steps such as telephoning the customer support. Figure 4 shows the working of sub-modules of User Verifier. Table 11 compares the other detection tools where it can be seen that no detection system is profiling the user behavior.

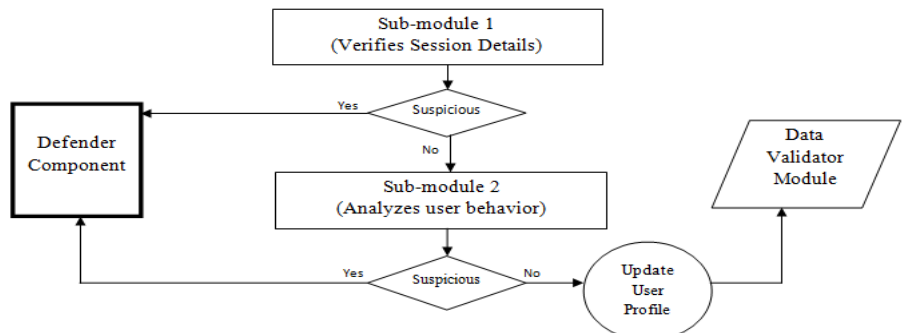

*Fig.4. Working of the User Verifier Module*

**Table 11** Comparison of *User Verifier* module features with others

| User Verifier Module | AppSensor | PHPIDS | ModSecurity | Shadow Daemon | Aqtronix WebKnight | Proposed IPS |
|---|---|---|---|---|---|---|
| Sub-module 1 | Yes | No | Yes | No | Yes | Yes |
| Sub-module 2 | No | No | No | No | No | Yes |

*8.2.5 Access controller:* Improper implementation of access control checks in the application may enable the adversaries to get unauthorized access to the privileged services (e.g. display credentials of all users) or resources (e.g. files, databases) [83]. This module primarily ensures that the requests to the operations and resources have been made in an authorized context. The Access Controller is comprised of three sub-modules, namely two-factor (2F) authenticator, role-based access controller (RBAC) and behaviour analyser that work in a sequence to recognize the unauthorized requests.

2F authentication is a security process that requires a user to provide the second piece of identifying information in addition to the password as the second step of verification[9]. It is a recommended practice for ensuring security to sensitive operations like an online transaction. The 2F authenticator sub-module inspects the requests to see whether it is for any critical service. The request for a sensitive operation is sent to the Defender where it prompts the user to enter the second authentication factor such as OTP. If the user supplies the valid credentials, the request is passed to the web server, otherwise rejected. The administrator may provide the list of sensitive operations to the module. If the request is not for a sensitive operation, it is passed to the second sub-module, RBAC.

Role-based access control is commonly used to manage the access control policies for web application [86]. In web application, each user is assigned a particular role (admin, member, visitor, etc.) which governs his privileges. The RBAC sub-module consists of a mapping of different roles assigned by the application to their respective set of permitted operations, and assess the requests based on some predefined rules for determining if the user has right to invoke the functionality. The sub-module is usually configured by the administrator. For large and complex web applications, the administrator may mistakenly forget to enlist some web operations in the policies. Therefore, the client request is verified against two conditions, first, if the requested operation is mentioned in the policies and second, if client's role is permitted to access the operation. If both the conditions are met, the request is finally considered as safe by all the modules of the Detector and so, the request is sent to the application for the processing. If the request fails to meet the second condition, the user is considered as unauthorized, and hence, the request is sent to the Defender. If the request violates the first condition i.e. the requested service is not even mentioned in the policies, it is passed to third sub-module for further verification.

The configuration of RBAC sub-module is mainly the responsibility of administrator, and therefore, is likely to be flawed. There can be services which the RBAC sub-module do not mention. The requests for such services are passed to the behavioral analyzer which is a machine learning-based sub-module. Various features such as the pattern of requested services and resources can help to build the role-based access behavioral profiles [81]. The submodule can also learn the resources which should not be directly accessible from the user to prevent insecure direct object reference vulnerability. The safe requests are sent to the web server, otherwise to the Defender component of the IPS. The behavioral analyzer may also assist the administrator to identify the flaws in the RBAC model as it points out the resources which are accessed by users but not mentioned in RBAC. The Figure 5 shows the working of the sub-modules and Table 12 provides the comparison.

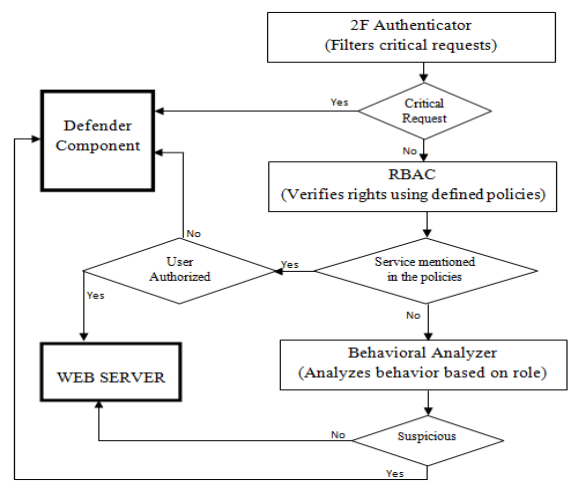

*Fig.5. Working of the Access Controller module*

---
[9] https://www.symantec.com/connect/blogs/guide-two-factor-authentication

**Table 12** Comparison of *Access Controller* module features with others

| Access Controller Module | AppSensor | PHPIDS | ModSecurity | Shadow Daemon | Aqtronix WebKnight | Proposed IPS |
|---|---|---|---|---|---|---|
| 2F Authenticator | No | No | No | No | No | Yes |
| RBAC | Yes | No | Yes | No | No | Yes |
| Behavioral Analyzer | No | No | No | No | No | Yes |

## 8.3. Defender

Once a suspicious request is detected, the next step is to decide how to deal with it. The Defender component takes appropriate countermeasures to prevent the website from adversarial operations in real time. The component is coupled with a set of response measures for translating the passive alert generated by the detector into automated actions. The response action is selected dynamically on the basis of various attack characteristics such as the nature of the attack, severity level, confidence level, etc. [67, 68]. Severity signifies the impact of damage to the targeted application and confidence level indicates the probability of the reported incident to be an attack. The detector component provides these values. For instance, in the case of session hijacking, the defender may forcefully log off the session and notify the account holder. An injection attack with high confidence and high severity (e.g., attempt to execute critical administrator operations) may cause the blocking of the user, whereas the attack with high confidence and low severity (e.g., just looking for vulnerability) would only cause the rejection of the current request. Moreover, for an intrusion with low confidence and low severity, the defender may log the incident and pass the request to the application for processing. Furthermore, upon recognizing the abnormal behavior with low confidence, the defender may decide to monitor the future requests over time and take appropriate action once the confidence level becomes higher. The possible actions that the defender component may initiate includes disabling the user account, disabling the compromised services, recording the incident, rejecting the current request, suspending the service, making the user logged out, suspending the user, blocking the IP address, suspending the IP address, etc. It is to be noted that the component will also update the black-list of Connection Verifier component to suspend or block the attacker, compromised services or resources. The Table 13 shows which of the detection tools is offering the functionalities of the Defender Component. Since PHPIDS and Aqtronix WebKnight operate in passive mode, they are mainly concerned with logging the records.

**Table 13** Comparison of *Defender* component features with others

| IPS Component | AppSensor | PHPIDS | ModSecurity | Shadow Daemon | Aqtronix WebKnight | Proposed IPS |
|---|---|---|---|---|---|---|
| Defender | Yes | No | Yes | Yes | No | Yes |

## 8.4. Response Controller

Response Controller intercepts the output to be sent to the client by the server to prevent any disclosure of critical and sensitive information, and also to keep the record of the data stored by the application in HTTP response. This component has three modules, namely Transformer, Marker and Encryptor. The following section briefly discusses the function of each module.

8.4.1 Transformer: This module performs two operations. First, it serves alternate content if there is any leaked information in the output and second, it embeds secret details (e.g., an anti-csrf token) in the response body. Application weakness such as improper server configurations and handling of error messages may reveal various sensitive information in the response body like technical details of the server, directory structure, etc [85]. The module replaces such information with some generic content. The module may use the pre-defined patterns for identifying the leaked information. Next, the module also appends an anti-csrf token to the HTTP response either as a form parameter or as a query string in the links to prevent session management related attacks. The token is later verified by the User Verifier module of the Detector component when the data is sent back to the server upon subsequent request.

8.4.2 Marker: This module assists the Detector component in encountering the parameter manipulation attacks. Its job is to generate the list that contains the records of the parameters that have been used by the application to store data on the client side. The Marker creates the hash [74] of such parameter values, form a list and supply it to the detector. The hash value assists in verifying the integrity of data content sent through insecure channels. When the data is submitted back, the Data Validator re-computes the hash of those values and compare with the hashes mentioned in the list (see section 8.2.3). The parameters used by the application to store values may be observed while training of the Data Validator module. During the training phase, the module can learn about the parameters which are initialized by the server side programs, and their values remain the same when submitted back to the server. The administrator may also provide the list of parameters to be marked.

8.4.3 Encryptor: Its main job is to encrypt the traffic for HTTPS communication. As discussed, the proposed IPS also provides the SSL offloading functionality to deal with the HTTPS request. The pre-processor component of the IPS converts the HTTPS requests into HTTP format, and the Encryptor sub-module is assigned the task to convert HTTP response to HTTPS if required. The Encryptor module can also be configured to provide security to HTTP communication. In the case of HTTP response, sensitive details such as cookies may need to be prevented from being seen by the client. This sub-module, if configured, can encrypt these values as well. When the subsequent request from the client returns these values, the pre-processor component can decrypt them.

Table 14 compares the detection systems with respect to the Response Controller features. We observed that only ModSecurity is controlling the HTTP response. However, it requires stringent rule specifications to serve the purpose. On the other hand, the Marker module of proposed IPS that assists in dealing with parameter tampering attacks can be trained via machine learning.

**Table 14** Comparison of *Response Controller* component features with others

| Response Controller Module | AppSensor | PHPIDS | ModSecurity | Shadow Daemon | Aqtronix WebKnight | Proposed IPS |
|---|---|---|---|---|---|---|
| Encryptor | No | No | Yes | No | No | Yes |
| Marker | Yes | No | Yes | No | No | Yes |
| Transformer | No | No | Yes | No | No | Yes |

## 8.5. Logger

The component works with two modules, namely the Defender Logger and Response Controller Logger to log the operations performed by Defender and Response Controller components respectively. The Defender is responsible for taking counter-measures against the suspicious incidents. The Defender Logger maintains details of the suspicious request as well as the action performed by Defender in response to suspicious activity. It logs all the relevant information of the attack incident such as username, IP address, time, type of attack, severity, requested URL, etc. The processing of these log records will help the administrator in analyzing the major threats on the application, tracking the intrusion as well as an intruder and assessing the performance of Detector and Defender components.

Similarly, another module maintains the details of the operations performed by the Response controller on the content sent to the client. The Response Controller broadly applies three functions on the response data, i.e., replace the potentially leaked information with some non-harmful content, marks the fields which are set by application programs and encryption. The details of these operations must be logged for analyzing and improving the performance of the Response Controller. Log details will help to ensure no critical value is sent in the plain text. Likewise, logs will also assist in tracking whether the component marks all the parameters set by the server side programs. If it is found that some parameters have been missed out by the Response Controller, appropriate changes can be made in its respective module. Moreover, the log details may also provide insight into the weaknesses of the application which leak the information.

Table 15 presents the comparison between the systems. It can be noted that the existing systems are mainly focussed on logging the details of the suspicious requests. But the Logger component also tracks the actions performed by the system on the content to be sent to the client.

**Table 15** Comparison of *Logger* component features with others

| Logger Module | AppSensor | PHPIDS | ModSecurity | Shadow Daemon | Aqtronix WebKnight | Proposed IPS |
|---|---|---|---|---|---|---|
| Defender Logger | Yes | Yes | Yes | Yes | Yes | Yes |
| Response Controller Logger | No | No | No | No | No | No |

The complete framework of IPS is shown in Figure 6. The framework has been designed to ensure the high-level security to the application. Some of the distinguishing features of the IPS are summarized as under:

- The *Request Filter* module of the *Preprocessor* component is novel in terms of functionality.
- The *Detector* module is equipped with a new set of features such as validating all types of parameter category and maintaining the profile of user behavior and access behavior.
- *Response Controller Logger* module is also unique in the proposed framework that tracks the actions performed by the system on HTTP response content.
- ModSecurity is the only tool that provides the maximum of the proposed features, but it requires the administrator to specify malicious behavior in the form of rules, whereas there are some modules in proposed IPS which use machine learning techniques to learn it.

However, it should be noted that there are certain limitations of the proposed framework of IPS. It just strengthens the security but does not eliminate the need for all security practices. For instance, it does not replicate the security provided by HTTPS communication. It won't encrypt the sensitive information coming from the client side. Therefore, if the user credentials are submitted to the server using HTTP protocol, the details can be intercepted by an attacker. Similarly, the framework does not have any hand in dealing with the issue of integrity of the client-side data. Furthermore, it does not eliminate the need for server-side validations. Input is validated only from the perspective of attacks only. It does not implement the application logic for verifying the values such as the uniqueness of username, the validity of the email, etc. Password quality is another issue which is not enforced by the IPS and can endanger the user security if the application permits its users to create easy, predictable and weak passwords.

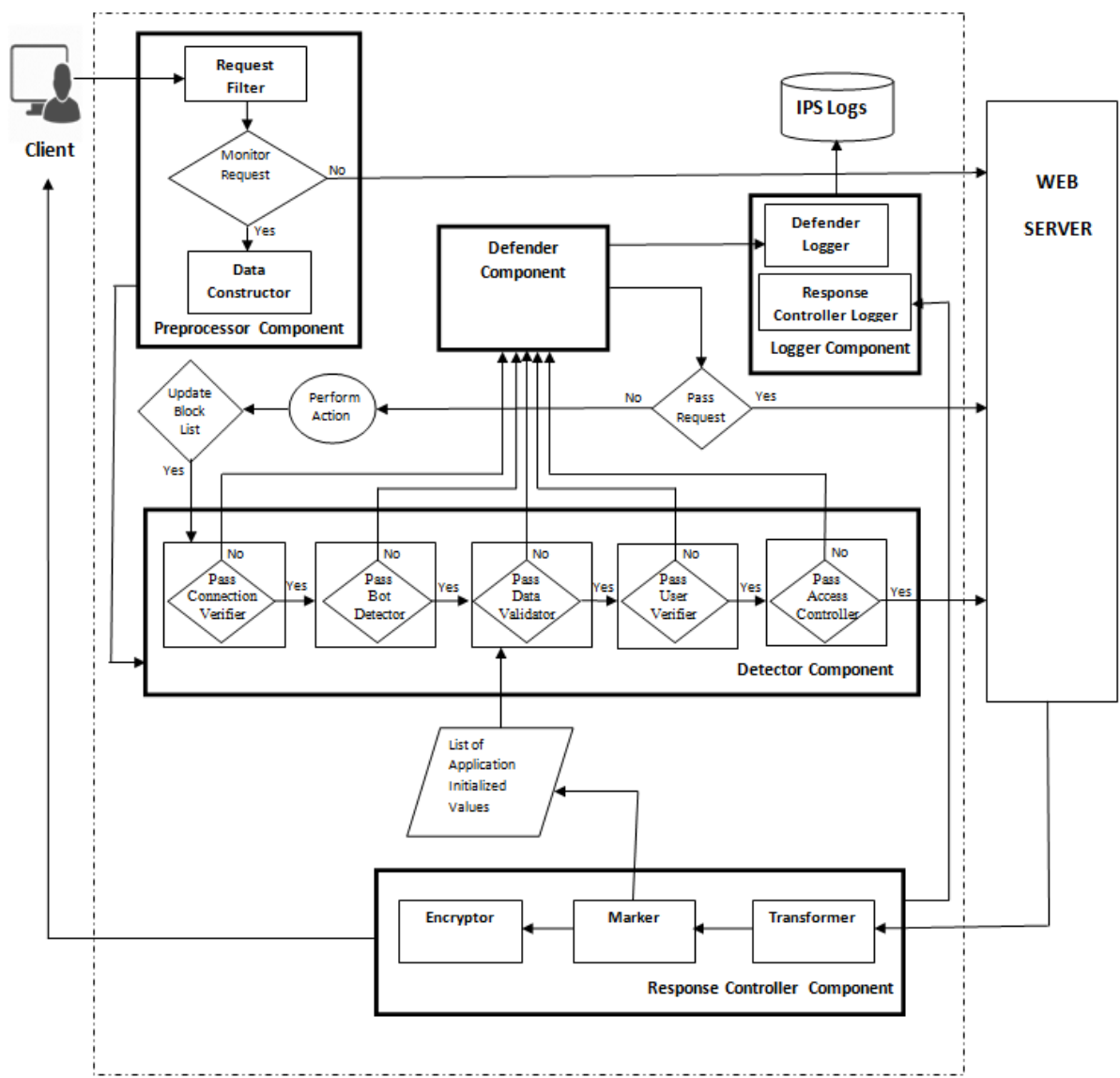

***Fig.6.*** *Framework of the proposed IPS*

## 9. Conclusion

Web application threats have become a prime concern for information security. IDS is one of the security mechanisms used to guard these applications against attacks. However, the methodology has been primarily used for monitoring the network-based attacks. Designing a suitable IDS to prevent web-based attacks still needs more focus by the interest groups. The paper provides all the essential information about the web IDS at a single place as a recipe to design an ideal intrusion detection system. In this paper, we have listed many particular characteristics of a web application which makes challenging for a developer to build an efficient detection system.  For example, HTTPS traffic makes the detector incapable of reading packet data if concerned SSL certificate is not provided. Web request parameter values are of various types such as text value, application value, format-specific value, etc., which requires the detector to use different validation approach for different types. Web applications facilitate multiple user interactions through sessions, and the IDS which do not track the user requests are not efficient enough to recognize attacks on session management and authorization policies. A web application can be implemented in various programming languages, and since NIDS inspects packets, it is independent of programming language, whereas HIDS are usually server-side language specific. Since software can also be used to generate HTTP requests, IDS should be capable of detecting bot requests. SIDS suits best for detecting scripts which are designed to automate web-based attacks, whereas the AIDS suits best for detecting scripts designed to mimic human behavior. Sound knowledge of all the discussed characteristics will assist in designing a robust architecture of the web IDS.

The paper also presents the evolution of the research work carried so far towards designing the intrusion detection systems for a web application. Several dimensions have been proposed to compare the existing detection systems on different parameters such as IDS type, IDS mode, data monitored type, response mechanism, etc. The overall comparative study has been summarized in terms of a graph which highlights various prominent facts about the current trends and the challenges in the domain.  The graph revealed that a few IDS are providing features such as output validation, stateful detection, prevention mechanism, dynamic learning and description of attacks. In fact, none of the systems mentioned in our survey guarantee to immune an application from all kinds of web attacks. Furthermore, a conceptual framework for an ideal web IPS has been proposed. The framework comprises of five components, namely *Preprocessor, Detector, Defender, Logger and Response Controller* The features of the framework have been compared with five well-known detection systems, namely *AppSensor, PHPIDS, ModSecurity, Shadow Daemon and AQTRONIX WebKnight*.

## 10.  Future Scope

The review conducted in the paper would prove to be extremely beneficial in identifying future avenues in the research domain of intrusion detection for a web application. We are on track to design a well-defined set of the guidelines for each component discussed in the framework that will assist developers and researchers to build an efficient IPS for the web application and subsequently implement them. The future of web IDS is a system that can learn custom attacks against an application, provide broader attack detection coverage, respond in real time and update itself over time.